\documentclass[prd,aps,twocolumn,tighten]{revtex4}
\usepackage{latexsym}
\begin{document}
\newcommand{\beq}{\begin{equation}}
\newcommand{\eeq}{\end{equation}}
\newcommand{\Prd}{Phys. Rev D}
\newcommand{\Prl}{Phys. Rev. Lett.}
\newcommand{\Plb}{Phys. Lett. B}
\newcommand{\Cqg}{Class. Quantum Grav.}
\newcommand{\Np}{Nuc. Phys.}

\title{The mass of the graviton and the cosmological constant}
\author{M. Novello and R. P. Neves}
\address{\mbox{}\\
Centro Brasileiro de Pesquisas F\'{\i}sicas,\\
Rua Dr.\ Xavier Sigaud 150, Urca 22290-180, Rio de Janeiro, RJ -- Brazil\\
E-mail: novello@cbpf.br}
\date{\today}

\begin{abstract}
We show that the graviton has a mass in an anti-de Sitter ($
\Lambda < 0$) background given by $m_{g}^{2}= -
\,\frac{2}{3}\Lambda.$ This is precisely the fine-tuning value
required for the perturbed gravitational field to mantain its two
degrees of freedom. \noindent
\end{abstract}

 \maketitle

\renewcommand{\thefootnote}{\arabic{footnote}}

\section{Introduction}

\bigskip

The mass of the graviton and the cosmological constant have been
seen, historically and conceptually, as two entirely different
quantities.  Nevertheless, we shall prove in this letter that
there is a very precise relation between these quantities,
expressed mathematically by the equation
$m_{g}^{2}=-\frac{2}{3}\Lambda.$

This result follows from the study of the propagation of spin-2
fields in curved spacetime. The construction of a coherent and
self-consistent theory for the spin-2 field in the general case of
curved riemannian spacetimes was, for a long time, thought to be
problematic \cite{Aragone-Deser-1, Aragone-Deser-2}. There were
difficulties related to the number of degrees of freedom,
causality of propagation and compatibility with the Einstein
equations. All these problems have a common origin in an apparent
ambiguity: since the covariant derivatives do not comute in the
curved spacetime, the generalization of the equation of motion for
the spin-2 field from flat to curved spacetime seems to involve an
arbitrary choice of the order of the derivatives.

In a recent paper \cite{Novello-Neves}, we showed that the
generalization of the Fierz variables \cite{Fierz-Pauli} to curved
spacetime avoids this ambiguity, leading to a unique, well-defined
equation of motion. The theory thus obtained is completely
self-consistent, without the need of any \textit{ad hoc}
hypothesis. The difficulties cited above turned out to be a
consequence of a bad choice for the equation of motion, and have
nothing to do with some anomaly in the physical world.

In this work, we will concentrate our analysis in the special case
of a massive spin-2 field propagating in de Sitter $(\Lambda>0)$
or anti-de Sitter $(\Lambda<0)$ spacetimes. We will show that,
when the mass of the field has the peculiar value
$m^{2}=-\frac{2}{3}\Lambda,$ which can be positive only in AdS,
the action has a gauge symmetry analogous to that of the massless
field in Minkowski spacetime, and, as a consequence of this
symmetry, the field has only two degrees of freedom. This
situation reproduces precisely what happens with the perturbations
of the Einstein equations in an AdS background: the perturbed
gravitons are shown to have a mass  with exactly this peculiar
value. We will also show that an "effective" mass, $M,$ can be
defined so that the field has two degrees of freedom only when $M$
vanishes. Otherwise, it has five degrees of freedom. Our
interpretation of $M$ as only an effective mass, and not the real
one (which we call $m$), is deeply rooted in the Fierz formalism.

A special value for the mass has also appeared in a recent paper
by Deser and Waldron \cite{Deser-Waldron}. They studied the
behavior of a spin-2 field in dS and AdS backgrounds and showed
that \ ''the cosmological constant $\Lambda $ and the mass
parameter define a phase plane in which partially massless gauge
invariant lines separate allowed regions from forbidden,
non-unitary ones''. The crucial result rests on the existence of a
spin-2 massive field that separates the two regions, the mass of
which is $m^{2}=\frac{2}{3} \,\Lambda .$ They argue that at this
particular value a novel symmetry appears that is responsible for
the elimination of the 0 helicity excitation. In their approach,
when the field has a mass with this value, it has four degrees of
freedom; it has two degrees of freedom only when $m=0.$ In the
interval $0<m^{2}< \frac{2}{3} \,\Lambda,$ the theory is unstable.
In the Minkowski and AdS geometries the theory is always stable
and the field has five degrees of freedom, unless it is massless.

Although some of their results agree with ours there are some
differences. First of all, we note that what they call the mass
should be compared with our effective mass, $M.$ We found that the
field has two degrees of freedom only when $M=0.$ The interval
$0<M^{2}< \frac{2}{3} \,\Lambda$ (when $\Lambda>0$) is forbidden,
since the real mass would have to be imaginary. In AdS, $M$ can
take any value (but the real mass should satisfy $m^{2}>
-\frac{2}{3} \,\Lambda$ for it not to be imaginary) and $M=m$ in
Minkowski. The main difference is that, as we have already pointed
out, the spin-2 field has either two or five degrees of freedom,
never four. The distinction between real and effective masses is
also a novelty. It makes it possible for us to state that the real
mass of the field does not vanish when it has two degrees of
freedom in the AdS geometry. These differences are certainly
related to the choice of the equation of motion for the spin-2
field in a curved spacetime. Our point of view is that the use of
the Fierz formalism leads to the best choice, as is extensively
discussed in \cite{Novello-Neves}.

It follows from all this that the real mass of the graviton is
related to a negative cosmological constant, even though its
effective mass vanishes. So, the anti-de Sitter spacetime replaces
the Minkowski spacetime as the basic vacuum solution of the
Einstein equations if the mass of the graviton is not null.

Finally, let us remark that it is quite an interesting property
that $ \frac{2}{3} \,\Lambda $ is a critical value in both
treatments. One should ask if very fine-tuning mass such as this
exists among all the spectrum of known particles.
 We shall prove in this paper that the particle which displays such
special mass is precisely what is called the graviton.

\bigskip

\section{\protect\bigskip The Fierz formalism in curved spacetime}

The Fierz formalism is an alternative way to describe a spin-2
field, characterized by the use of a three-index tensor
$F_{\alpha\beta\mu},$ which is anti-symmetric in the first pair of
indices and obeys the cyclic identity, that is
\begin{equation}
F_{\alpha \beta \mu }+F_{\beta \alpha \mu }=0, \label{2.1}
\end{equation}
\begin{equation}
F_{\alpha \beta \mu }+F_{\beta \mu \alpha }+F_{\mu \alpha \beta
}=0. \label{2.2}
\end{equation}
Since such an object has $20$ independent components, we must
impose an additional condition that eliminates $10$ of them,
otherwise it will describe two spin-2 fields. We set\footnote{We
define the symmetrization symbol by $A_{(\mu\nu)} \equiv
A_{\mu\nu} + A_{\nu\mu}.$}:

\begin{equation}
\stackrel{\ast }{F}{}^{\alpha (\mu \nu)}{}_{,\alpha }=0,
\label{2.3}
\end{equation}
in which the asterisk represents the dual operator.

 Condition (\ref{2.3}) implies that there exists a
symmetric second order tensor $\varphi _{\mu \nu }$ such that we
can write
\begin{equation}
2\,F_{\alpha \beta \mu }=\varphi _{\mu \alpha ,\beta }-\varphi
_{\mu \beta ,\alpha }+F_{\alpha }\,\eta _{\beta \mu }-F_{\beta
}\,\eta_{\alpha \mu }, \label{2.4}
\end{equation}
where $F_{\alpha }=F_{\alpha \beta \mu }\eta ^{\beta \mu }=\varphi
_{,\alpha }-\varphi _{\alpha }{}^{\beta }{}_{,\beta }.$

We call $F_{\alpha \beta \mu }$ a \textit{Fierz tensor}. It
satisfies the identity
\begin{equation}
F^{\alpha }{}_{(\mu \nu ),\alpha }\equiv
-2\,G^{(\ell)}{}_{\mu\nu}, \label{2222}
\end{equation}
where $G^{(\ell )}{}_{\mu\nu}$ is the linearized Einstein
operator, defined in terms of the symmetric tensor $\varphi
_{\mu\nu }$ by
\begin{equation}
G^{(\ell )}{}_{\mu \nu }\equiv \frac{1}{2}\left[ \Box
\,\varphi_{\mu\nu}-\varphi^{\alpha }{}_{(\mu ,\nu
)\,,\alpha}+\varphi _{,\mu\nu}-\eta _{\mu\nu}\,\left( \Box \varphi
-\varphi ^{\alpha\beta}{}_{,\alpha\beta}\right) \right].
\label{2.7}
\end{equation}

This tensor\footnote{We changed the notation used in our previous
quoted paper introducing a factor $\frac{1}{2}$ in the definition
of $G^{(\ell)}{}_{\mu\nu}$ that makes it better suited for
applications in the general relativity context.} has the important
property of being divergence-free.

The generalization of the Fierz formalism to a curved riemannian
background is made through the use of the minimal coupling
principle, replacing $\eta_{\mu \nu }$ by $g_{\mu \nu }$ and the
simple derivatives by covariant ones in equation (\ref{2.4}):
\begin{equation}
2F_{\alpha \beta \mu }=\varphi _{\mu \alpha ;\beta }-\varphi _{\mu
\beta;\alpha }+F_{\alpha }g_{\beta \mu }-F_{\beta }g_{\alpha \mu}.
\label{2.8}
\end{equation}

As a direct consequence of the fact that the covariant derivatives
do not comute in a curved spacetime, condition (\ref{2.3}) becomes
\begin{equation}
\stackrel{\ast }{F}^{\alpha }{}_{(\mu \nu )\,;\alpha }=\frac{1}{2}
R_{\alpha (\mu \nu )\beta }^{\text{ \ \ \ \ \ \ }\ast
}{}\,\,\varphi^{\alpha \beta }. \label{2.9}
\end{equation}
In the special cases of de Sitter or anti-de Sitter spacetimes, we
recover the result $\stackrel{\ast}{F}^{\alpha
(\mu\nu)}{}_{;\label{2.10}\alpha}=0.$ The role of
$G^{(\ell)}{}_{\mu\nu}$ is now played by the tensor
$\widehat{G}_{\mu\nu},$ defined so that
\begin{equation}
F^{\alpha}{}_{(\mu\nu);\alpha }\equiv -2\widehat{G}_{\mu\nu}.
\label{299}
\end{equation}
A straightforward calculation shows that
\begin{equation}
\widehat{G}_{\mu\nu} \equiv \,\frac{1}{2} \left[
G^{(a)}{}_{\mu\nu}+G^{(b)}{}_{\mu\nu} \right] \label{2.12}
\end{equation}
where
\begin{equation}
G^{(a)}{}_{\mu \nu }\equiv \frac{1}{2}\left[ \Box \,\varphi _{\mu
\nu }-\varphi _{\alpha (\mu ;\nu )}{}^{;\alpha }+\varphi _{;\mu
\nu}-g_{\mu \nu }\, \left( \Box \varphi -\varphi ^{\alpha \beta
}{}_{;\alpha\beta }\right) \right]
\label{2.13}
\end{equation}
and
\begin{equation}
G^{(b)}{}_{\mu \nu }\equiv \frac{1}{2}\left[ \Box \,\varphi _{\mu
\nu}-\varphi _{\alpha (\mu }{}^{;\alpha }{}_{;\nu
)}+\varphi_{;\mu\nu}-g_{\mu \nu }\,\left( \Box \varphi
-\varphi^{\alpha \beta }{}_{;\alpha\beta }\right) \right] .
\label{2.14}
\end{equation}

\bigskip

\section{The massive spin-2 field in a de Sitter background}

\subsection{The equation of motion}

In the Minkowski spacetime, the equation of motion for a massive
spin-2 field is
\begin{equation}
G^{(\ell )}{}_{\mu \nu } + \frac{1}{2}m^{2}\,\left( \varphi
_{\mu\nu }-\varphi \,\eta _{\mu \nu }\right) =0.
 \label{3.1.1}
\end{equation}

Taking the divergence of this equation and inserting it back in
its trace, we find that we must have
\begin{equation}
\varphi =0, \text{ \ \ \ \ }\varphi _{\mu }^{\text{ \ \
}\nu}{}_{,\nu }=0,
\label{3.1.2}
\end{equation}
so that the field has five degrees of freedom.

When we pass from flat to curved spacetime, the generalization of
equation (\ref{3.1.1}) is not trivial. Thanks to the fact that the
covariant derivatives no longer comute, we could make different
choices for the order in which they appear. We could use either
$G^{(a)}{}_{\mu \nu }$ or $G^{(b)}{}_{\mu \nu }$ to replace
$G^{(\ell )}{}_{\mu \nu },$ or even any combination of both. Since
the fundamental work of Aragone and Deser \cite{Aragone-Deser-1},
it has become a tradition the use of $G^{(a)}{}_{\mu \nu }.$
However, there is no solid motivation for this particular choice.
In a recent work, we showed that the extension of the Fierz
formalism to curved spacetime provides a well motivated choice,
once it is completely unambiguous. Using Fierz variables, we
arrive at the following equation of motion:

\begin{equation}
F^{\alpha}{}_ {(\mu \nu);\alpha } - m^{2}\, \left(\varphi _{\mu
\nu }-\varphi\,g_{\mu \nu }\right) =0 \label{333}
\end{equation}
or, equivalently:
\begin{equation}
\widehat{G}_{\mu \nu } + \frac{1}{2}m^{2}\,\left( \varphi _{\mu
\nu }-\varphi\,g_{\mu \nu }\right) =0.
 \label{3.1.3}
\end{equation}
Its divergence yields
\begin{equation}
Z^{\mu } - m^{2}\,F^{\mu }=0, \label{3.1.4}
\end{equation}
where we define $Z^{\mu }\equiv -F^{\alpha \,(\mu
\nu)}{}_{;\alpha\nu }=2\widehat{G}^{\mu \nu }{}_{;\nu }.$ It is
not difficult to show that $Z^{\mu}$ is given by
\begin{equation}
Z^{\mu }=R^{\alpha \beta \lambda \mu }\,F_{\alpha \beta
\lambda}-\frac{1}{2}\,\left( R_{\varepsilon }{}^{[\lambda
}\varphi^{\mu]\varepsilon }\right) _{;\lambda }.
\label{3.1.6}
\end{equation}

In the Minkowski spacetime, $Z^{\mu}$ vanishes identically, but
this is no longer true in the curved spacetime. When (\ref{3.1.4})
is not an identity, we impose the following conditions, in order
to have a consistent system of equations:
\begin{equation} \varphi =0,
\label{3.1.73}
\end{equation}
\begin{equation} F^{\mu}=0,
\label{3.1.72}
\end{equation}
\begin{equation}
 Z^{\mu}=0.
\label{3.1.7}
\end{equation}

The vanishing of $Z^{\mu}$ should be seen as four conditions to be
satisfied by the metric tensor $g_{\mu \nu }.$

When the background is a de Sitter spacetime, in which $R_{\alpha
\beta \mu\nu}=\frac{\Lambda }{3}\left( g_{\alpha \mu
}g_{\beta\nu}-g_{\alpha \nu}g_{\beta \mu }\right) ,$ equation
(\ref{3.1.6}) becomes
\begin{equation}
Z^{\mu }=-\frac{2}{3} \Lambda \,F^{\mu },
\label{3.1.8}
\end{equation}
and we see that the conditions (\ref{3.1.73}), (\ref{3.1.72}) and
(\ref{3.1.7}) are self-consistent.
\subsection{The fine-tuning mass}

\bigskip
Equation (\ref{3.1.3}) is derived from the action

\begin{equation}
S=\frac{1}{4}\int \sqrt{-g}\left[ A-B - \frac{m^{2}}{2}\left(
\varphi _{\mu\nu }\varphi ^{\mu \nu }-\varphi ^{2}\right)
\right]d^{4}x, \label{3.2.1}
\end{equation}
where\bigskip\ $A\equiv F_{\alpha \mu \nu
}\hspace{0.5mm}F^{\alpha\mu \nu }$ and $B\equiv F_{\mu
}\hspace{0.5mm}F^{\mu }.$

In the flat Minkowski background (where $g_{\mu\nu} =
\eta_{\mu\nu}$) the massless field equation is invariant under the
transformation $\varphi _{\mu \nu }\rightarrow \varphi _{\mu \nu
}+\delta \varphi _{\mu \nu }$ given by
\begin{equation}
\delta \varphi _{\mu \nu }=\xi _{\mu \label{transf};\nu }+\xi
_{\nu ;\mu }.
 \label{3.2.4}
\end{equation}
In a curved spacetime the same transformation causes the action
(\ref{3.2.1}) to change according to
\begin{equation}
\delta S=\frac{1}{2}\int \sqrt{-g}\left( Z^{\mu } -
m^{2}F^{\mu}\right) {}\xi_{\mu }d^{4}x.
 \label{3.2.5,1}
\end{equation}
In general, this action will not be invariant under such a
transformation. However when the background is a de Sitter
spacetime, equations (\ref{3.1.8}) and (\ref{3.2.5,1}) tell us
that in the very particular case when the mass of the spin-2 field
has the special value $m^{2}= - \, \frac{2}{3}\Lambda ,$ the
action recovers a gauge invariance\footnote{Note that the gauge
invariance of the massless field in Minkowski spacetime is
correctly contained in the limit $\Lambda \rightarrow 0.$ }. The
surprising feature of this result is that the action invariance,
in a de Sitter background, does not occur for the massless field,
but for the field that satisfies $m^{2}= - \, \frac{2}{3}\Lambda
.$ For any other value of the mass, the conditions (\ref
{3.1.73}),(\ref {3.1.72}) and (\ref {3.1.7}) must be imposed.

 In section IV we shall prove that the particle which displays such
special mass is precisely what we could call the graviton.

\subsection{The effective mass}

When the field is massless, the equation of motion reduces to
$\widehat{G}^{\mu \nu }=0.$ Taking the divergence of this
equation, in a de Sitter spacetime, we see from equation
(\ref{3.1.8}) and the definition of $Z^{\mu}$ that we must have
$F^{\mu}=0.$ Besides, in this particular geometry the following
identity holds:
\begin{equation}
\widehat{G}^{\mu \nu
}{}_{;\mu\nu}-\frac{\Lambda}{3}\widehat{G}^{\mu \nu }
g_{\mu\nu}=0. \label{3.3.1}
\end{equation}

This eliminates a further degree of freedom, which we can
conveniently choose to be given by $\varphi=0.$ So, the conditions
(\ref{3.1.73}), (\ref{3.1.72}) and (\ref{3.1.7}) are satisfied,
and we can see that the field has five degrees of freedom.

This is a rather curious result. Indeed, the massless field has
the properties we would expect from a massive field, while the
special features that we usually associate to the massless field
occurs when the mass has the fine-tuning value $m_{g}^{2}=
-\frac{2}{3}\Lambda.$ This suggests that we define an effective
mass for the spin-2 field in a de Sitter spacetime, given by
\begin{equation}
M^{2}=m^{2} + \frac{2}{3}\Lambda. \label{3.3.2}
\end{equation}
Then, the field has two degrees of freedom only when $M = 0.$ For
any other value of $M,$ it has five degrees of freedom.

It is important to notice that, when $\Lambda>0,$ the interval $0<
M^{2}< \frac{2}{3} \,\Lambda$ is forbidden, since the real mass,
$m,$ cannot be imaginary. Similarly, the interval $0< m^{2}<
-\frac{2}{3} \,\Lambda$ renders an imaginary effective mass when
$\Lambda<0.$ It seems, thus, that $m^{2}=-\frac{2}{3}\Lambda$ is a
lower bound for the real mass in the AdS geometry, corresponding
to the situation in which the field has only two degrees of
freedom. The real mass is not restricted in dS, but the effective
mass is. Inversely, it is the effective mass that is not
restricted in AdS. Note that $M=0$ is forbidden in the case in
which $\Lambda>0,$ so that the field cannot have two degrees of
freedom in dS.

\section{The fine-tuning mass of the graviton}

\bigskip
Let us perturb the Einstein equations of motion (with the
cosmological term),
\begin{equation}
G_{\mu \nu }+\Lambda \text{ }g_{\mu \nu }=0, \label{4.1}
\end{equation}
around the de Sitter solution $\stackrel{\text{o}}{g}_{\mu\nu.}$
Using the notation $\delta g_{\mu \nu }=\varphi _{\mu \nu },$ the
equation for the perturbed field is
\begin{equation}
\delta G_{\mu \nu }+\Lambda \varphi _{\mu \nu }=0,
\label{4.3}
\end{equation}
where $\delta G_{\mu \nu }$ is the perturbation of the Einstein
tensor. A direct calculation yields
\begin{equation}
\delta G_{\mu \nu }=G^{(a)}{}_{\mu \nu} +A_{\mu \nu },
\end{equation}
\label{4.4}
where
\begin{equation}
A_{\mu \nu }=\frac{1}{2}\left( \stackrel{\text{o}}{g}_{\mu \nu }\stackrel{%
\text{o}}{R}^{\alpha \beta}\varphi _{\alpha
\beta}-\stackrel{\text{o}}{R}\varphi _{\mu \nu }\right) .
\label{4.5}
\end{equation}
Manipulating equations (\ref{2.12}), (\ref{2.13}) and
(\ref{2.14}), we arrive at
\begin{equation}
G^{(a)}{}_{\mu \nu }=\widehat{G}_{\mu
\nu}-\frac{1}{2}R_{\mu\alpha\nu \beta }\,\varphi ^{\alpha
\beta}+\frac{1}{4}\,R_{\alpha (\mu}\varphi _{\nu )}{}^{\alpha }.
\label{4.6}
\end{equation}
In the de Sitter background, equation (\ref{4.3}) becomes
\begin{equation}
\widehat{G}_{\mu \nu }-\frac{\Lambda }{3}\left( \varphi _{\mu
\nu}-\varphi \stackrel{\text{o}}{g}_{\mu \nu }\right) =0.
\label{4.10}
\end{equation}

Comparing with equation (\ref{3.1.3}), one recognizes that the
graviton has a mass $m_{g}^{2}=-\frac{2}{3}\Lambda.$ In the
anti-de Sitter spacetime, according to our previous discussion,
this is the real value for the graviton mass, which in turn is
associated with a null effective mass, so that the field has only
two degrees of freedom, as we showed above.

\bigskip
\section{Conclusion}
The use of the Fierz formalism leads to a well-defined equation of
motion for the spin-2 field, avoiding all the arbitrariness
present in other approaches. Using this equation, we analysed the
case of a massive spin-2 field propagating in a de Sitter or in
anti-de Sitter background, and showed that, when the mass has the
special value $m_{g}^{2}=- \,\frac{2}{3}\Lambda $ - the
\textit{fine-tuning mass}-  which can be positive only in anti-de
Sitter, the action has a gauge symmetry analogous to the one
present in Minkowski spacetime, and the field has only two degrees
of freedom. For any other mass value, including $m = 0,$ the field
has five degrees of freedom. This led us to define an effective
mass $M$ such that the field has two degrees of freedom whenever
it vanishes and five otherwise.

We were also able to show that a perturbation of the Einstein
equations (with the cosmological constant term), in a de Sitter or
anti-de Sitter background, produces exactly the equation we
obtained from the Fierz formalism for the massive spin-2 field.
The graviton mass
is related to the cosmological constant by precisely the
fine-tuning value described above - a relation that makes sense
only when $\Lambda < 0.$ So, although not strictly massless, the
graviton, in an anti-de Sitter background, keeps the same number
of degrees of freedom it has in the Minkowski spacetime, since its
effective mass vanishes. Inverting the argument, we may say that
if the graviton has a real mass that is not null, then the
cosmological constant should be negative.

\section{Acknowledgement}

This work was partially supported by the Brazilian agency Conselho
Nacional de Desenvolvimento Cient\'{\i}fico e Tecnol\'ogico
(CNPq).

\end{document}